\documentclass[11pt,onecolumn,draft]{IEEEtran}
\usepackage{times,cite,lineno}
\usepackage{amsmath,amssymb,amsfonts,latexsym,epsf,enumerate,cases,algorithm,algorithmic,hhline,multirow,fancyhdr,subfigure}

\begin{document}

\title{\huge{Regularized Zero-Forcing Interference Alignment for the Two-Cell MIMO Interfering Broadcast Channel}}

\author{\large{Joonwoo~Shin$^{*}$ and Jaekyun~Moon,~\IEEEmembership{Fellow,~IEEE}}

\thanks{This work was supported in part by the IT R\&D program of MKE/KEIT
(KI0038765,Development of B4G Mobile Communication Technologies
for Smart Mobile Services). The authors are with the School of
EECS, Korea Advanced Institute of Science and Technology (KAIST),
373-1,
Guseong-dong, Yuseong-gu, Daejeon, 305-701, Republic of Korea (e-mail: joonoos@etri.re.kr, jmoon@kaist.edu).}
}
%
%

\maketitle

\begin{abstract} In this
paper, we propose transceiver design strategies for the two-cell
multiple-input multiple-output (MIMO) interfering broadcast
channel where inter-cell interference (ICI) exists in addition to
inter-user interference (IUI). We first formulate the generalized
zero-forcing interference alignment (ZF-IA) method based on the
alignment of IUI and ICI in multi-dimensional subspace. We then
devise a minimum weighted-mean-square-error (WMSE) method based on
``regularizing" the precoders and decoders of the generalized
ZF-IA scheme. In contrast to the existing
weighted-sum-rate-maximizing transceiver, our method does not
require an iterative calculation of the optimal weights. Because
of this, the proposed scheme, while not designed specifically to
maximize the sum rate, is computationally efficient and achieves a
faster convergence compared to the known weighted-sum-rate
maximizing scheme. Through analysis and simulation, we show the
effectiveness of the proposed regularized ZF-IA scheme.

\end{abstract}


\IEEEpeerreviewmaketitle

\section{Introduction}

\PARstart{M}ulti-cell and multi-user downlink transmission schemes
such as network MIMO and coordinated multi-point (CoMP)
transmission and reception methods have received a great deal of
attention for being able to boost the system performance with base
station (BS) cooperation. As a practical scenario of the
multi-cell and multi-user downlink transmission, one may consider
the heterogeneous networks, e.g., macro-pico or macro-femto
cellular networks where the dominant interference can be much
stronger than the residual interferences from adjacent cells. This
scenario can be modelled as a two-cell interfering broadcast
channel (IBC).

To improve communication over the two-cell IBC, various MIMO
transmission strategies that combine the spectral efficiency of
MIMO spatial division multiple access and the interference
mitigation capability of BS cooperation have been investigated
\cite{Shi_WMMSE_WSR, Shin_WMMSE_WSR,
ILee_MIMO_2Cell_WiCOM,ShinWJ_WiCoM_2011,Suh_DL}. An iterative
weighted-sum-rate-maximizing transceiver design method for the
multi-cell MIMO IBCs has been proposed
\cite{Shi_WMMSE_WSR,Shin_WMMSE_WSR}. An analytical expression for
the degree of freedom (DoF) for the two-cell MIMO IBC has been
provided in \cite{ILee_MIMO_2Cell_WiCOM}. However, the
corresponding achievable DoF is distinctly lower than the trivial
outer-bound on DoF of \cite{Jafar_DoF}. To improve the DoF, the
authors of \cite{ShinWJ_WiCoM_2011,Suh_DL} have introduced
modified interference alignment (IA) methods which reduce the
interference dimension by aligning ICI or IUI. The IA condition of
\cite{ShinWJ_WiCoM_2011}, however, has been developed for the
limited user configuration of two users per cell. In
\cite{Suh_DL}, a zero-forcing IA (ZF-IA) method for the $K$-user
per cell case has been proposed.
 It is well known that the original MIMO IA method of
\cite{Jafar_IT}, which has been developed for the MIMO
interference channel, is sub-optimal at any finite SNR regime
despite of its ability to achieve the DoF. Given the sub-optimality of IA
in the interference channel, it is reasonable to expect the suboptimality
of ZF-IA at finite SNRs for the IBC.

We accordingly propose a new IA scheme based on ZF-IA for the
two-cell MIMO IBC. To proceed, we first generalize the ZF-IA
method of \cite{Suh_DL} from \emph{single} stream transmission to
\emph{multiple} stream transmission for each communication link.
Then, to improve the sum-rate at finite SNRs, we propose a method
of ``regularizing" the ZF-IA scheme based on the WMSE criterion.
Through analysis and numerical simulation, we verify that the
proposed regularized ZF-IA scheme indeed improves on the
generalized ZF-IA method and outperforms the existing weighted
sum-rate-maximizing method if the number of iterations for
transceiver filter computation is limited.

The following notations are used. We employ upper case boldface
letters for matrices and lower case boldface letters for vectors.
For any general matrix ${\bf{X}}$, ${\bf{X}}^*$, ${\bf{X}}^H$,
$\text{Tr}({\bf{X}})$, $\text{det}({\bf{X}})$,
  and $\text{SVD}(${\bf{X}}$)$ denote the conjugate, the Hermitian
transpose, the trace, the determinant, and the singular value
decomposition, respectively.  The symbol ${\bf{I}}_n$ denotes an
identity matrix of size $n$.

\section{System Model}

We consider the two-cell MIMO interfering broadcast channel. The
$m$-th base station $\textsf{B}_m$ equipped with $M$ antennas
supports $K$ users $\lbrace\textsf{D}_{mk}\rbrace$ in the
corresponding cell, and each user has $N$ antennas $\big(m\in
(1,2), k\in (1,\cdots,K)\big)$. Denoting ${\bf{y}}^{\lbrack m,k
\rbrack}$ as the signal vector received by the $k$-th user in the
$m$-th cell $\textsf{D}_{mk}$, the two-cell MIMO interfering
broadcast channel is mathematically described as
\begin{align}
{\bf{y}}^{\lbrack m,k \rbrack}=&{{\bf{H}}^{\lbrack
m,k\rbrack}_{m}}{\bf{T}}^{\lbrack m,k \rbrack}{\bf{s}}^{\lbrack
m,k \rbrack}+{{\bf{H}}^{\lbrack m,k \rbrack}_{m}}\sum_{i\ne
k}^{K}{\bf{T}}^{\lbrack m,i\rbrack}{\bf{s}}^{\lbrack
m,i\rbrack}\nonumber\\& +{{\bf{H}}^{\lbrack
m,k\rbrack}_{\overline{m}}}\sum_{i=1}^{K}{\bf{T}}^{\lbrack
\overline{m},i\rbrack}{\bf{s}}^{\lbrack\overline{m},i\rbrack}+{\bf{n}}^{\lbrack
m,k\rbrack} \label{y_SigM}
\end{align}
where ${\bf{T}}^{\lbrack m, k\rbrack}\in{\mathcal{C}^{M\times
L_s}}$ is the precoding matrix for $\textsf{D}_{mk}$,
${\bf{s}}^{\lbrack m, k\rbrack}\in{\mathcal{C}^{L_s\times 1}}$
stands for the signal vector of length $L_s$ transmitted for
$\textsf{D}_{mk}$, ${\bf{n}}^{\lbrack m,k\rbrack}$ is the additive
Gaussian noise at  $\textsf{D}_{mk}$ with
$\mathcal{CN}(0,\sigma_n^2)$ and ${\bf{H}}^{\lbrack
m,k\rbrack}_{\overline{m}} \in \mathcal{C}^{N\times M}$ is the
channel matrix from $\textsf{B}_{\overline{m}}$ to
$\textsf{D}_{mk}$; here we define $\overline{1}=2$ and
$\overline{2}=1$. It is assumed that the channel elements are
independent identically distributed (i.i.d.) complex Gaussian
random variables with zero mean and unit variance and
$\mathbb{E}\lbrack {\bf{s}}^{\lbrack m,k\rbrack}{\bf{s}}^{\lbrack
m,k\rbrack H}\rbrack={\bf{I}}_{L_s}$. The transmit precoder at
$\textsf{B}_{m}$ satisfies the power constraint
$\sum_{k}\text{Tr}({\bf{T}}^{\lbrack m,k\rbrack
}{{\bf{T}}^{\lbrack m,k\rbrack}}^{H})\le P_m$, where $P_m$ is the
maximum transmit power of  $\textsf{B}_{m}$. The estimated output
vector at $\textsf{D}_{mk}$ is obtained with the receive filter
${\bf{U}}^{\lbrack m,k\rbrack}\in \mathcal{C}^{M\times L_s}$ as
$\hat{\bf{s}}^{\lbrack m,k\rbrack}= {{\bf{U}}^{\lbrack
m,k\rbrack}}^{H}{\bf{y}}^{\lbrack m,k\rbrack}$.

\section{Two-cell zero-forcing interference alignment}

In this section, we present a generalized zero-forcing IA (ZF-IA)
method in the two-cell MIMO interfering broadcast channel. First,
we briefly review the existing ZF-IA scheme. Then, we describe
generalized ZF-IA for multiple stream transmission for each
link. This generalized ZF-IA will serve as a basis of the
regularized ZF-IA scheme which will be described in Section
\ref{Sect:prop}.

\subsection{Review of the ZF-IA
method}\label{Sect:ReviewDLIA}

To achieve $\frac{K}{K+1}$ DoF-per-cell\footnote{Compared to the
DoF definition in \cite{Jafar_IT}, the notion of DoF-per-cell is based on
normalization of the DoF by the dimensionality.} without BS
cooperation, the transmit precoders are written as
${\bf{T}}^{\lbrack m,k\rbrack}={\bf{P}}{\bf{v}}^{\lbrack
m,k\rbrack}$ where ${\bf{P}}\in \mathcal{C}^{M\times N_p}$ is
introduced for each BS to spread $N_p$ streams over
$M$-dimensional transmit antenna resource $(M>N_p)$ and
${\bf{v}}^{\lbrack m,k\rbrack}$ \cite{Suh_DL}. The ZF-IA method is
available in the symmetric antenna configuration \cite{Suh_DL};
from this point on we focus on symmetric cases, i.e. $M=N$.

The ZF-IA scheme of \cite{Suh_DL} assumes a single stream
reception with each receiver filter $\lbrace {\bf{u}}^{\lbrack
m,k\rbrack}\rbrace$. The receive filter output of
$\textsf{D}_{mk}$ is written as (\ref{s_sigM_Suh}) in the below,
\begin{figure*}
\begin{align}
\hat{{s}}^{\lbrack m,k\rbrack}=&{{\bf{u}}^{\lbrack
m,k\rbrack}}^{H} {\overline{\bf{H}}^{\lbrack
m,k\rbrack}_{m}}{\bf{v}}^{\lbrack m,k\rbrack}{{s}}^{\lbrack
m,k\rbrack}+{{\bf{u}}^{\lbrack
m,k\rbrack}}^{H}{\overline{\bf{H}}^{\lbrack
m,k\rbrack}_{m}}\sum_{i\ne k}^{K}{\bf{v}}^{\lbrack
m,i\rbrack}{{s}}^{\lbrack m,i\rbrack}+{{\bf{u}}^{\lbrack
m,k\rbrack}}^{H}{\overline{\bf{H}}^{\lbrack
m,k\rbrack}_{\overline{m}}}\sum_{i=1}^{K}{\bf{v}}^{\lbrack\overline{m},i\rbrack}{{s}}^{\lbrack\overline{m},i\rbrack}+{{\bf{u}}^{\lbrack
m,k\rbrack}}^{H}{\bf{n}}^{\lbrack m,k\rbrack} \label{s_sigM_Suh}
\end{align}
\hrulefill
\end{figure*} where ${\overline{\bf{H}}^{\lbrack
m,k\rbrack}_{\overline{m}}}\triangleq {{\bf{H}}^{\lbrack
m,k\rbrack}_{\overline{m}}}{\bf{P}}$. To null out the ICI, the
third term on the right hand side of (\ref{s_sigM_Suh}),
${{\bf{u}}^{\lbrack m,k\rbrack}}\in{\mathcal{C}^{M\times 1}}$,
lies in the null space of ${\overline{\bf{H}}^{\lbrack
m,k\rbrack}_{\overline{m}}}$. To guarantee the existence of these
receive filters $\lbrace {\bf{u}}^{\lbrack m,k\rbrack}\rbrace$,
the dimension of the spreading matrix ${\bf{P}}$ should be
$(K+1)\times K$, i.e. $M=K+1$ and $N_{p}=K$. The remaining IUI is
cancelled with a transmit channel inversion method
\cite{Hochwald_CI}.  From the ICI nulling process
${{\bf{u}}^{\lbrack m,k\rbrack}}^{H}{\overline{\bf{H}}^{\lbrack
m,k\rbrack}_{\overline{m}}}={\bf{0}}^{T}$ and IUI cancellations
${{\bf{u}}^{\lbrack m,k\rbrack}}^{H}{\overline{\bf{H}}^{\lbrack
m,k\rbrack}_{m}}{\bf{v}}^{\lbrack m,i\rbrack}=0$ $(i\ne k)$,  it
is easily verified that both ICI and IUI are aligned in the null
space of ${\bf{u}}^{\lbrack m,k\rbrack}$.

\subsection{Generalized ZF-IA}\label{Sect:generalized}

Although the ZF-IA scheme achieves $\frac{K}{K+1}$ DoF-per-cell,
only a \emph{single} stream transmission is allowed at each user
node. To transmit $L_s$ $(L_s>1)$ streams at each user node, we
propose a generalized ZF-IA transceiver design method. At first,
for the spreading matrix ${\bf{P}}$, to guarantee the existence of
null space of ${\overline{\bf{H}}^{\lbrack
m,k\rbrack}_{\overline{m}}}$ with $\text{rank}=L_s$, we pick an
arbitrary $K(L_s+1)\times KL_s$ full rank matrix whose columns are
orthonormal to each other, i.e.,
${\bf{P}}^{H}{\bf{P}}={\bf{I}}_{N_p}$. Then, the received signal
(\ref{y_SigM}) is rewritten as
\begin{align}
{\bf{y}}^{\lbrack m, k\rbrack}=&{\overline{\bf{H}}^{\lbrack
m,k\rbrack}_{m}}{\bf{V}}^{\lbrack m,k\rbrack}{\bf{s}}^{\lbrack
m,k\rbrack}+{\overline{\bf{H}}^{\lbrack m,k\rbrack}_{m}}\sum_{i\ne
k}^{K}{\bf{V}}^{\lbrack m,i\rbrack}{\bf{s}}^{\lbrack
m,i\rbrack}\nonumber \\ &+{\overline{\bf{H}}^{\lbrack
m,k\rbrack}_{\overline{m}}}\sum_{i=1}^{K}{\bf{V}}^{\lbrack
\overline{m},i\rbrack}{\bf{s}}^{\lbrack
\overline{m},i\rbrack}+{\bf{n}}^{\lbrack m,k\rbrack}
\label{y_SigM_ZFIA}
\end{align}
where ${\bf{V}}^{\lbrack m,k \rbrack}\in{\mathcal{C}^{N_p\times
L_s}}$.
 To cancel out the ICI, the front end of the receiver filter
${{\bar{\bf{U}}}^{\lbrack m,k\rbrack}} \in \mathcal{C}^{M\times
L_s}$ is chosen from the null space of
${\overline{\bf{H}}}^{\lbrack m,k\rbrack}_{\overline{m}}$, which
can be obtained as
\begin{equation}
\text{SVD}({\overline{\bf{H}}}^{\lbrack
m,k\rbrack}_{\overline{m}})=\lbrack
{{{\tilde{\bar{\bf{U}}}}}^{\lbrack m,k\rbrack}},
{{\bar{\bf{U}}}^{\lbrack m,k\rbrack}}\rbrack
{\bar{\bf{\Sigma}}}^{\lbrack m,k\rbrack}{\bar{\bf{V}}}^{\lbrack
m,k\rbrack H} \nonumber
\end{equation}
Now  $\textsf{B}_{m}$ performs block diagonalization (BD) to
eliminate IUI, the BD precoder ${\bar{\bf{V}}}^{\lbrack
m,k\rbrack}$ is identified as
\begin{equation}
\text{SVD}({\bf{H}}_{\text{C}}^{\lbrack
m,k\rbrack})=\bar{\bf{U}}_{\text{C}}^{\lbrack
m\rbrack}{\bar{\bf{\Sigma}}}_{\text{C}}^{\lbrack
m\rbrack}\lbrack{\tilde{\bar{\bf{V}}}}^{\lbrack
m,k\rbrack},{\bar{\bf{V}}}^{\lbrack
m,k\rbrack}\rbrack^{H}\label{V_bar_BD_intraIC},
\end{equation}
where ${\bf{H}}_{\text{C}}^{\lbrack m,k\rbrack}= \Big{\lbrack}
{\bf{\Omega}}_{m}^{\lbrack m,1 \rbrack H},
\cdots,{\bf{\Omega}}_{m}^{\lbrack m,k-1 \rbrack
H},{\bf{\Omega}}_{m}^{\lbrack m,k+1 \rbrack H},\cdots
\Big{\rbrack}^{H}$ and ${\bf{\Omega}}_{\overline{m}}^{\lbrack m,k
\rbrack}\triangleq {{\bar{\bf{U}}}^{\lbrack m,k\rbrack
H}}{\overline{\bf{H}}^{\lbrack m,k\rbrack}_{\overline{m}}}$.

Assume the final ZF-IA transceivers are
${\bf{T}}_{\text{GZF-IA}}^{\lbrack
m,k\rbrack}={\bf{P}}\bar{\bf{V}}^{\lbrack
m,k\rbrack}\hat{\bf{V}}^{\lbrack m,k\rbrack}{\bf{\Phi}}^{\lbrack
m,k\rbrack\frac{1}{2}}$ and ${\bf{U}}_{\text{GZF-IA}}^{\lbrack
m,k\rbrack}=\bar{\bf{U}}^{\lbrack m,k\rbrack}\hat{\bf{U}}^{\lbrack
m,k\rbrack}$; then the estimated signal is written as
\begin{equation}
\hat{\bf{s}}^{\lbrack m,k\rbrack}=\hat{\bf{U}}^{\lbrack m,k\rbrack
H}{\bf{H}}^{\lbrack m,k\rbrack}_{\text{eff}}\hat{\bf{V}}^{\lbrack
m,k\rbrack}{\bf{\Phi}}^{\lbrack m,k \rbrack\frac{1}{2}
}{\bf{s}}^{\lbrack m,k \rbrack}+{\hat{\bf{n}}}^{\lbrack
m,k\rbrack}, \label{s_SigM_BD_IA}
\end{equation}
where the effective channel ${\bf{H}}^{\lbrack
m,k\rbrack}_{\text{eff}}\in{\mathcal{C}^{L_s\times L_s}}$ and the
effective noise ${\hat{\bf{n}}}^{\lbrack
m,k\rbrack}\in{\mathcal{C}^{L_s\times 1}}$ are defined by
${\bf{H}}^{\lbrack m,k\rbrack}_{\text{eff}}=\bar{\bf{U}}^{\lbrack
m,k\rbrack H}{\overline{\bf{H}}}^{\lbrack
m,k\rbrack}_{m}\bar{\bf{V}}^{\lbrack m,k\rbrack}$ and
${\hat{\bf{n}}}^{\lbrack m,k\rbrack}= \hat{\bf{U}}^{\lbrack
m,k\rbrack H}\bar{\bf{U}}^{\lbrack m,k\rbrack
H}{{\bf{n}}}^{\lbrack m,k\rbrack}$, respectively. The other
transmit-receive matrices  $ \hat{\bf{V}}^{\lbrack m,k\rbrack}$
and $ \hat{\bf{U}}^{\lbrack m,k\rbrack}$ are identified by channel
diagonalization with $\text{SVD}\big({\bf{H}}^{\lbrack
m,k\rbrack}_{\text{eff}}\big)= {\hat{\bf{U}}^{\lbrack
m,k\rbrack}}{\hat{\bf{\Sigma}}^{\lbrack
m,k\rbrack}}{\hat{\bf{V}}^{\lbrack m,k\rbrack H}}$. Note that
because ${\hat{\bf{U}}^{\lbrack m,k\rbrack}}$ and
${\bar{\bf{U}}^{\lbrack m,k\rbrack}}$ are composed of orthonormal
columns, $\mathbb{E}(\hat{\bf{n}}^{\lbrack
m,k\rbrack}\hat{\bf{n}}^{\lbrack m,k\rbrack
H})=\sigma_n^2{\bf{I}}_{L_s}$. Then the information rate of
$\textsf{D}_{mk}$ can be computed as
\begin{equation}
\text{R}^{\lbrack m,k\rbrack}_{\text{ZF-IA}}=\text{log}\lbrace
\text{det}
({\bf{I}}_{L_s}+{\sigma_n^{-2}}{{\hat{\bf{\Sigma}}}^{\lbrack
m,k\rbrack 2 }{\bf{\Phi}}^{\lbrack m,k\rbrack }})\rbrace.
\end{equation}
Because this scheme causes no ICI, the sum rate over the $m$-th
cell $\sum_{k=1}^{K} \text{R}^{\lbrack m,k\rbrack}_{\text{ZF-IA}}$
is independent of the power allocation at
$\textsf{B}_{\overline{m}}$. Thus, the sum-rate-maximizing power
allocation problem
\begin{align}
&\max_{\lbrace {\bf{\Phi}}^{\lbrack m,k\rbrack}\rbrace}
\sum_{m=1}^{2}\sum_{k=1}^{K} \text{R}^{\lbrack
m,k\rbrack}_{\text{ZF-IA}} \quad \text{subject
to}\sum_{k=1}^{K}\text{Tr}({\bf{\Phi}}^{\lbrack m,k\rbrack})\le
P_{m}, \forall m \nonumber
\end{align} is divided into the following \emph{individual-cell} sum-rate-maximizing
problem (in which the optimal power allocation matrix $\lbrace
{\bf{\Phi}}^{\lbrack m,k\rbrack}\rbrace$ is calculated with the
water-filling solution)
\begin{align}
&\max_{\lbrace {\bf{\Phi}}^{\lbrack m,k\rbrack}\rbrace}
\sum_{k=1}^{K} \text{R}^{\lbrack m,k\rbrack}_{\text{ZF-IA}} \quad
\text{subject to}\sum_{k=1}^{K}\text{Tr}({\bf{\Phi}}^{\lbrack
m,k\rbrack})\le P_{m} \label{Opt_pwr_alloc}
\end{align}
where the power constraint
$\text{Tr}({\bf{T}}_{\text{GZF-IA}}^{\lbrack
m,k\rbrack}{\bf{T}}_{\text{GZF-IA}}^{\lbrack m,k\rbrack
H})=\text{Tr}({\bf{\Phi}}^{\lbrack m,k\rbrack})$ is obtained using
${\bf{P}}^{H}{\bf{P}}={\bf{I}}_{N_p}$. Let us call this scheme
\emph{generalized} ZF-IA (GZF-IA). Note that the proposed GZF-IA
scheme still preserves $\frac{K}{K+1}$
DoF-per-cell\footnote{DoF-per-cell is
$\frac{2KL_s}{2(K+1)L_s}=\frac{K}{K+1}$.} and is implemented
without BS cooperation.

\section{Proposed regularized ZF-IA
method}\label{Sect:prop}

Due to the inherent limitations of ZF schemes, the original IA
method of \cite{Jafar_IT} is distinctly sub-optimal in the
low-to-mid SNR regime. We surmise that GZF-IA is also suboptimal.
We propose a regularized GZF-IA algorithm which regularizes the
precoders and decoders of the GZF-IA scheme in an effort to
improve upon the sum rate performance of GZF-IA.

\subsection{Transceiver design}
To achieve regularization, the proposed scheme minimizes the
weighted MSE defined as
\begin{align}
&\min \sum_{m=1}^{2} \sum_{k=1}^{K} \mathbb{E} \lbrace
|{\bf{\Lambda}}^{\lbrack m,k\rbrack}{\bf{s}}^{\lbrack
m,k\rbrack}-\hat{\bf{s}}^{\lbrack m,k\rbrack} |^2\rbrace \nonumber
\\ &\text{subject to}\sum_{k=1}^{K}\text{Tr}({\bf{T}}^{\lbrack
m,k\rbrack}{\bf{T}}^{\lbrack m,k\rbrack H})\le P_{m}, \forall m
\label{WMSE}
\end{align}
where ${\bf{\Lambda}}^{\lbrack m,k\rbrack}$ is introduced to
improve the sum-rate performance by preventing weaker subchannels
from being assigned more power. Accordingly,
${\bf{\Lambda}}^{\lbrack m,k\rbrack}$ is chosen as the effective
channel gain matrix to $\textsf{D}_{mk}$, ${\bf{\Lambda}}^{\lbrack
m,k\rbrack}={\bf{U}}^{\lbrack m,k\rbrack
H}_{\text{GZF-IA}}{\bf{H}}^{\lbrack
m,k\rbrack}_{m}{\bf{T}}^{\lbrack m,k\rbrack}_{\text{GZF-IA}}$.
Then the Lagrangian function of (\ref{WMSE}) is formed as
(\ref{Lag}) in the below,
\begin{figure*}
\begin{align}
\mathcal{L}=  \sum_{m=1}^{2} \sum_{k=1}^{K} \text{Tr}&
\Big{\lbrace} {\bf{\Lambda}}^{\lbrack m,k\rbrack 2} -
{\bf{U}}^{\lbrack m,k\rbrack H} {\overline{\bf{H}}}^{\lbrack
m,k\rbrack}_{m} {\bf{V}}^{\lbrack
m,k\rbrack}{\bf{\Lambda}}^{\lbrack m,k\rbrack H} -
{\bf{\Lambda}}^{\lbrack m,k\rbrack } {{\bf{V}}}^{\lbrack
m,k\rbrack H } {\overline{\bf{H}}}^{\lbrack m,k\rbrack
H}_{m}{\bf{U}}^{\lbrack m,k\rbrack} + \sigma_n^2
{{\bf{U}}}^{\lbrack m,k\rbrack H}{{\bf{U}}}^{\lbrack
m,k\rbrack}\nonumber\\ & + \sum_{n=1}^{2}\sum_{i=1}^{K}
{\bf{U}}^{\lbrack m,k\rbrack H}{\overline{\bf{H}}}^{\lbrack
m,k\rbrack }_{n} {\bf{V}}^{\lbrack n,i\rbrack}{\bf{V}}^{\lbrack
n,i\rbrack H}{\overline{\bf{H}}}^{\lbrack m,k\rbrack
H}_{n}{\bf{U}}^{\lbrack m,k\rbrack}
\Big{\rbrace}+\sum_{m=1}^{2}\mu_{m}\Big{(}\sum_{k=1}^{K}\text{Tr}({\bf{V}}^{\lbrack
m,k\rbrack}{\bf{V}}^{\lbrack m,k\rbrack H})-P_m\Big{)} \label{Lag}
\end{align}
\hrulefill
\end{figure*}
 where $\lbrace\mu_m \rbrace$ is the
Lagrangian multiplier and the transmit power at $\textsf{B}_{m}$
is given by $\text{Tr}({\bf{T}}^{\lbrack
m,k\rbrack}{\bf{T}}^{\lbrack m,k\rbrack
H})=\text{Tr}({\bf{V}}^{\lbrack m,k\rbrack}{\bf{V}}^{\lbrack
m,k\rbrack H})$ using ${\bf{P}}^{H}{\bf{P}}={\bf{I}}_{N_p}$.
Because the transceiver matrix $\lbrace{\bf{V}}^{\lbrack
m,k\rbrack}\rbrace$ and $\lbrace{\bf{U}}^{\lbrack
m,k\rbrack}\rbrace$ are inter-related, it is difficult to optimize
simultaneously. Thus, we rely on an alternating optimization
method which iteratively finds local optimal solutions. First, we
design the optimal precoder assuming the receive filters are
given. From $\nabla_{{\bf{V}}^{\lbrack m,k
\rbrack*}}\mathcal{L}={\bf{0}}$, the precoder for
$\textsf{D}_{mk}$ is derived as:
\begin{align}
{{\bf{V}}}^{\lbrack m,k\rbrack }=  &\Big{(} \sum_{n=1}^{2}\sum_{i=
1}^{K}  {\bf{\Xi}}^{\lbrack n,i\rbrack}_{m}+\mu_m {\bf{I}}_{N_p}
\Big{)}^{-1}{\overline{\bf{H}}}^{\lbrack m,k\rbrack
H}_{m}{\bf{U}}^{\lbrack m,k\rbrack} {\bf{\Lambda}}^{\lbrack
m,k\rbrack} \label{V_mk}
\end{align}
 where  ${\bf{\Xi}}^{\lbrack
n,i\rbrack}_{m}\triangleq{\overline{\bf{H}}}^{\lbrack n,i\rbrack
H}_{m}{\bf{U}}^{\lbrack n,i \rbrack}{\bf{U}}^{\lbrack n,i \rbrack
H} {\overline{\bf{H}}}^{\lbrack n,i\rbrack}_{m}$. Since the $m$-th
BS transmit power, $\sum_{k=1}^{K} \text{Tr}({\bf{V}}^{\lbrack
m,k\rbrack}{\bf{V}}^{\lbrack m,k\rbrack H})$, is a monotonically
decreasing function with respect to $\mu_m$ (the proof is omitted
due to the space limitation), $\mu_m$ can be efficiently solved to
satisfy the power constraint by a bisection method.

Next, we derive the receive filter $\lbrace {\bf{U}}^{\lbrack m,k
\rbrack}\rbrace$ with the given precoders $\lbrace
{\bf{V}}^{\lbrack m,k \rbrack}\rbrace$. The optimal receive filter
for $\textsf{D}_{mk}$ is simply derived with
$\nabla_{{\bf{U}}^{\lbrack m,k \rbrack*}}\mathcal{L}={\bf{0}}$ and
is given by:
\begin{align}
{{\bf{U}}}^{\lbrack m,k\rbrack}=  &\Big{\lbrace}
\sum_{n=1}^{2}\sum_{i= 1}^{K} {\bf{\Psi}}^{\lbrack
m,k\rbrack}_{\lbrack n,i\rbrack}+\sigma_n^2 {\bf{I}}_{M}
\Big{\rbrace}^{-1}{\overline{\bf{H}}}^{\lbrack m,k\rbrack}_{m}
{\bf{V}}^{\lbrack m,k\rbrack} {\bf{\Lambda}}^{\lbrack m,k\rbrack
H}\label{U_mk}
\end{align}
where ${\bf{\Psi}}^{\lbrack m,k\rbrack}_{\lbrack
n,i\rbrack}\triangleq{\overline{\bf{H}}}^{\lbrack m,k\rbrack}_{n}
{\bf{V}}^{\lbrack n,i\rbrack}{\bf{V}}^{\lbrack n,i\rbrack
H}{\overline{\bf{H}}}^{\lbrack m,k\rbrack H}_{n}$. Since the
transceivers in (\ref{V_mk}) and (\ref{U_mk}) are inter-dependent,
the algorithm shown in the table below is used to find the
optimal transceivers.
\begin{algorithm}
\caption{Obtaining optimal regularized ZF-IA transceivers}
\label{Alg:proposed}
\begin{algorithmic}
\STATE{Initialize ${\bf{U}}^{\lbrack
m,k\rbrack}={\bf{U}}_{\text{GZF-IA}}^{\lbrack m,k\rbrack}$ and
compute the MSE weight ${\bf{\Lambda}}^{\lbrack m,k \rbrack}$,
$\forall m,k$.}

\STATE{Step l: Compute $\lbrace {\bf{V}}^{\lbrack
m,k\rbrack}\rbrace$ using (\ref{V_mk}).}

\STATE{Step 2: Compute $\lbrace {\bf{U}}^{\lbrack
m,k\rbrack}\rbrace$ using (\ref{U_mk}).}

\STATE{Step 3: Go back to Step 2 until convergence.}
\end{algorithmic}
\end{algorithm}
This algorithm is provable convergent at least to a local minimum.

We note that even though the MSE weights $\lbrace
{\bf{\Lambda}}^{\lbrack m,k \rbrack} \rbrace$ of the proposed
regularized ZF-IA algorithm is not optimum in sense of the sum
rate, they are obtained non-iteratively with the GZF-IA method,
which is near-optimum in the high SNR region. In the following, we
discuss the advantages of \emph{one-shot} calculation of
the MSE weights.

\section{Discussion: Computational complexity and prerequisite information exchange}

Here we analyze computational complexity and the amount of
prerequisite information of the proposed regularized ZF-IA
(RZF-IA) method. For comparison, we also analyze those of the
weighted-sum-rate-maximizing method (called `max-WSR method') of
\cite{Shi_WMMSE_WSR}.

\subsection{Computational complexity}

\begin{figure}
\centering \mbox{\subfigure[Computational
complexity]{\leavevmode\epsfxsize=0.5\textwidth
\epsffile{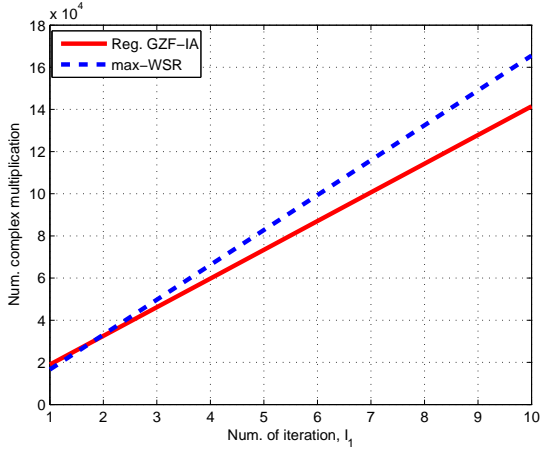}}\quad
\subfigure[Prerequisite information to
$\textsf{B}_m$]{\leavevmode\epsfxsize=0.5\textwidth
\epsffile{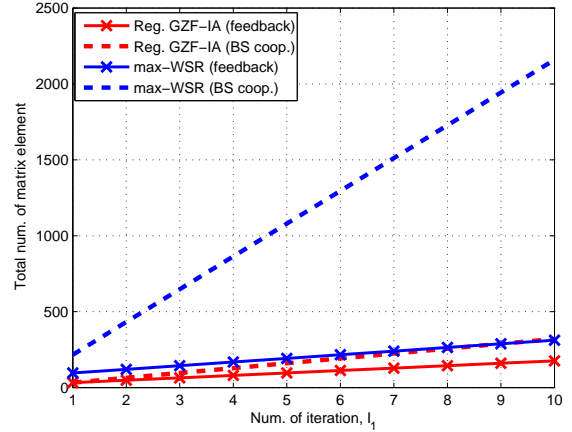}}}
\caption{Computation complexity and feedback/BS cooperation
resources required versus number of iterations, $I_1$}
\label{FIG:Analysis}
\end{figure}

We consider the number of complex multiplications as a complexity
measure. Fig.\ref{FIG:Analysis} (a) illustrates the computational
complexity for $K=2$, $M=6$, $L_s=2$, $N_p(=M-KL_s)=2$ and $I_2$
(the number of iterations for bisection) $=10$. In each iteration,
both RZF-IA and max-WSR schemes calculate the transmit and
receiver filters. The max-WSR scheme additionally includes
MSE-weight updating in the iteration loop, whereas the MSE weights
of RZF-IA are calculated in a non-iterative manner. Therefore, as
the number of iterations $I_1$ increases, the computational
efficiency of the RZF-IA method becomes relatively higher.

\subsection{Prerequisite information exchange}

To find the weighted-MSE-minimizing transmit precoders, each BS
requires prerequisite information through feedback and BS
cooperations.  Due to the one-shot calculation of the MSE weights
in RZF-IA, only the effective channels  $ {\bf{U}}^{\lbrack
m,k\rbrack H}{\bf{H}}_{m}^{\lbrack m,k\rbrack}{\bf{P}} \in
\mathcal{C}^{L_s\times N_p},\forall k$ are fedback iteratively for
updating $\lbrace {\bf{V}}^{\lbrack m,k \rbrack} \rbrace$.
However, the max-WSR method requires the channel information and
receiver filter coefficients separately to update the transmit
filters as well as MSE weights.  For the same reason, RZF-IA
requires a smaller amount of resources for BS cooperations.
Fig.\ref{FIG:Analysis}(b) clearly shows that the RZF-IA scheme is
advantageous in terms of the amount of prerequisite information.
Note that unlike GZF-IA which can be implemented without BS
cooperation, both RZF-IA and max-WSR require BS cooperation.
Nevertheless, considering that BS cooperation will be part of
future wireless communication standards \cite{3GPP}, the overhead
associated with BS cooperation of both methods seems reasonable.

\section{Numerical Results}

\begin{figure}[t]
\begin{center}
\leavevmode\epsfxsize=0.5\textwidth
\epsffile{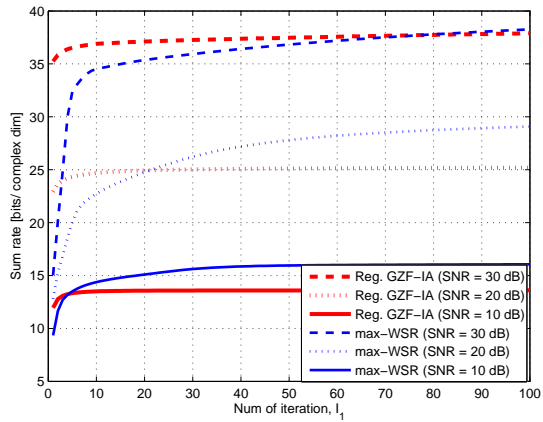}
\caption{{\small{Convergence of RZF-IA and max-WSR methods}}}
\label{FIG:Convergence}
\end{center}
\end{figure}

\begin{figure}[t]
\begin{center}
\leavevmode\epsfxsize=0.5\textwidth
\epsffile{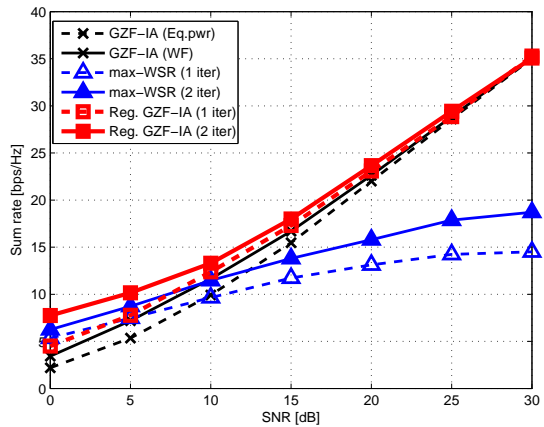} \caption{{\small{Sum
rate performance at small number of iterations}}}
\label{FIG:SR}
\end{center}
\end{figure}

This section evaluates the sum rate performance of various
transmission strategies over two-cell MIMO interfering broadcast
channels. For the simulation results, we set $M=6$, $K=2$,
$L_s=2$, $P_m=P,\forall m$. The SNR is defined as
$\frac{P}{\sigma_n^2}$. Also, we assume that the elements of the
channel matrix are i.i.d. complex Gaussian with zero mean and unit
variance. Fig. \ref{FIG:Convergence} illustrates the convergence
behavior of the RZF-IA method and max-WSR method. This plot shows
that while RZF-IA is not as good as max-WSR as a large number of iterations is allowed,
especially at low SNRs, the former algorithm converges faster than
the latter method. In fact, at a small number of iteration, RZF-IA performs better than max-WSR.

Fig. \ref{FIG:SR} shows the sum rate
performance at a small number of iterations $I_1=1,2$.
Specifically, at $I_1=2$, due to the fast convergence, RZF-IA
indeed shows better performance than max-WSR. We also confirm that RZF-IA
enhances the performance of GZF-IA. At a
sufficient number of iterations, e.g., at $I_1=100$,
the RZF-IA scheme shows a significant degradation, especially when SNR is not very large,
compared to max-WSR due to the sub-optimality of the MSE weights,
a price paid for reduced computational complexity and prerequisite
information.

\section{Conclusion}

In this paper, we have investigated generalized ZF-IA in the
two-cell MIMO interfering broadcast channel and subsequently proposed
regularized ZF-IA methods to improve its sum rate performance. To
execute the regularization process efficiently, we have utilized
the WMSE metric whose weight terms are computed from the effective
channel gain of the generalized ZF-IA scheme.  With these weights,
the regularized ZF-IA method iteratively calculates the transceivers.
Unlike the existing max-WSR method where weights
are found with iterations, the weights of the regularized ZF-IA
scheme are obtained \emph{non-iteratively} from the generalized
ZF-IA method. Overall, the proposed regularized ZF-IA scheme consumes less resources
and converges faster.
Through analysis and numerical simulation, the
effectiveness of the regularized ZF-IA scheme has been confirmed.

%

\ifCLASSOPTIONcaptionsoff
  \newpage
\fi
\bibliographystyle{IEEEtran}
\bibliography{IEEEabrv,References}


\end{document}